\documentclass{article}

\usepackage{arxiv}

\usepackage[utf8]{inputenc} 
\usepackage[T1]{fontenc}    
\usepackage{hyperref}       
\usepackage{url}            
\usepackage{booktabs}       
\usepackage{amsfonts}       
\usepackage{nicefrac}       
\usepackage{microtype}      
\usepackage{lipsum}
\usepackage{graphicx}
\graphicspath{ {./images/} }

\usepackage[round]{natbib}

\usepackage{amsmath,amsfonts,bm}

\def\eqref#1{equation~\ref{#1}}

\def\1{\bm{1}}

\DeclareMathAlphabet{\mathsfit}{\encodingdefault}{\sfdefault}{m}{sl}
\SetMathAlphabet{\mathsfit}{bold}{\encodingdefault}{\sfdefault}{bx}{n}

\usepackage{hyperref}
\usepackage{url}

\usepackage{amsmath,amssymb,amsfonts,amsthm}
\usepackage{algorithmic}
\usepackage{graphicx}
\usepackage{bbm}
\usepackage{textcomp}
\usepackage{xcolor}
\usepackage{esvect}
\usepackage{mathtools}
\usepackage{physics}
\usepackage{stmaryrd}
\usepackage{subcaption}
\usepackage{enumitem}
\usepackage{circledsteps}

\usepackage{tikz}
\usetikzlibrary{3d, shapes, arrows.meta, positioning}

\def\BibTeX{{\rm B\kern-.05em{\sc i\kern-.025em b}\kern-.08em
    T\kern-.1667em\lower.7ex\hbox{E}\kern-.125emX}}

\newtheorem{definition}{Definition}

\newtheorem{observation}{Observation}

\title{Layers of a City: Network-Based Insights into San Diego's Transportation Ecosystem}

\author{
  Matthew Chan \\
  ECE, UC San Diego\\
  \texttt{mac010@ucsd.edu} \\
   \And
 Steve Sharp \\
  ECE, UC San Diego\\
  \texttt{ssharp@ucsd.edu} \\
  \And
 Jiajian Zhu \\
  ECE, UC San Diego\\
  \texttt{jiz184@ucsd.edu} \\
  \And
 Raman Ebrahimi \\
  ECE, UC San Diego\\
  \texttt{raman@ucsd.edu} \\
}

\ifodd 1
\newcommand{\com}[1]{{\color{red}\textbf{Parinaz's Comment}: #1}}
\newcommand{\comr}[1]{{\color{orange}\textbf{Raman's Comment}: #1}}
\newcommand{\resp}[1]{{\color{cyan}\textbf{Response}: #1}} 
\else

\newcommand{\com}[1]{}
\newcommand{\comr}[1]{}
\newcommand{\resp}[1]{}
\fi

\begin{document}
\maketitle
\begin{abstract}
    Analyzing the structure and function of urban transportation networks is critical for enhancing mobility, equity, and resilience. This paper leverages network science to conduct a multi-modal analysis of San Diego's transportation system. We construct a multi-layer graph using data from OpenStreetMap (OSM) and the San Diego Metropolitan Transit System (MTS), representing driving, walking, and public transit layers. By integrating thousands of Points of Interest (POIs), we analyze network accessibility, structure, and resilience through centrality measures, community detection, and a proposed metric for walkability.

    Our analysis reveals a system defined by a stark core-periphery divide. We find that while the urban core is well-integrated, 30.3\% of POIs are isolated from public transit within a walkable distance, indicating significant equity gaps in suburban and rural access. Centrality analysis highlights the driving network's over-reliance on critical freeways as bottlenecks, suggesting low network resilience, while confirming that San Diego is not a broadly walkable city. Furthermore, community detection demonstrates that transportation mode dictates the scale of mobility, producing compact, local clusters for walking and broad, regional clusters for driving. Collectively, this work provides a comprehensive framework for diagnosing urban mobility systems, offering quantitative insights that can inform targeted interventions to improve transportation equity and infrastructure resilience in San Diego.
\end{abstract}

\section{Introduction}

Modern urban planning faces the immense challenge of understanding and optimizing complex, multi-modal transportation systems \cite{litman2017introduction}. Cities are intricate networks where residents travel via numerous modes, including walking, driving, and public transportation \cite{alessandretti2023multimodal}. Improving our comprehension of how these different networks are structured and how they interact is essential for enhancing urban mobility and guiding effective city planning. By dissecting the organizational fabric of these systems, we can identify key connections, constraints, and opportunities to create more efficient and accessible urban environments.

To tackle this challenge, network science offers a powerful framework\cite{alessandretti2023multimodal, ding2019application}. Graphs serve as an exceptional tool for modeling and measuring associations within large, complex datasets, such as those representing a city's infrastructure \cite{li2024opencity, tian2024analyzing, robson2021structure, phua2024fostering, zhang2025metacity, schoonenberg2019hetero}. This project leverages graph-based analysis to explore real-world transportation data with an emphasis on practical utility. The OpenStreetMaps (OSM) dataset \cite{OpenStreetMap}, a rich, collated source of international map and infrastructure data, provides the foundation for this analysis, allowing for the creation of intuitive and visual graphs that function as maps. The OSMnx Python library \cite{boeing2017osmnx} , built upon the NetworkX library \cite{hagberg2008exploring}, is used to parse this dataset and construct the graph models for analysis.

This paper addresses the central research question: \emph{How can a multi-layered graph analysis of San Diego's transportation networks reveal the functional differences between transport modalities and identify opportunities for improved urban mobility and accessibility?} To answer this, the study focuses on the San Diego area as a familiar local context. The core of the analysis involves creating and comparing separate graph layers for different transportation modes, initially contrasting driving and walking pathways. This approach is then expanded to incorporate additional layers, including Points of Interest (POIs) from OSM and detailed public transit data from the San Diego Open Dataset portal, which provides route and stop information not fully available in the main dataset.

The investigation unfolds in two distinct stages. After the dataset description, the paper establishes a baseline understanding by characterizing the fundamental properties of individual transportation layers—driving and walking—through a analysis of communities formed using network science methods. The latter employs community detection techniques to identify densely connected, functional transport zones within different modalities and in combined, multi-layer graphs, revealing localized transport clusters and their underlying structure. Second, it constructs and analyzes a multi-layer network that integrates POIs with public transportation stops and routes to assess the accessibility of key urban locations. Then, continuing the more sophisticated analysis, we extract various verifiable facts and observations using network attributes analysis. Collectively, these analyses aim to model the structural and functional organization of mobility in San Diego, leveraging network theory to provide actionable insights into the city's transportation system. Lastly, we would like to emphasize that even though this study was done on the available public data from San Diego, it can be applied to any other city and would achieve similar \emph{actionable} results. 

\subsection{Related Work}
In this section, we review some of the previous works that our work is related to, and is inspired by. 

Our study of San Diego's transportation network is informed by a rich body of literature that uses graph theory to model, analyze, and optimize urban systems. The foundational decision in this field involves how to represent a city's street network as a graph. In their chapter, \citep{agryzkov2017different} discuss the two primary methods: the primal approach, where intersections become nodes, and the dual approach, where streets are nodes. They emphasize that this initial modeling choice is not trivial, as each representation highlights different aspects of the urban structure and significantly influences subsequent analysis. Building on this, \citep{courtat2011centrality} introduce a novel framework that moves from a purely topological representation to a geometrical one, arguing that traditional models fail to capture the continuous nature of urban environments. Their use of color-coded ``centrality maps'' allows for an intuitive visual analysis of a city's structure, a diagnostic approach relevant to our goal of identifying key corridors and potential congestion points in San Diego.

Once a model is established, complex network theory offers a powerful paradigm for analysis. \citep{ding2019application} offer a comprehensive review of this field, categorizing models and applications such as congestion analysis and resilience, and providing a broad theoretical foundation for applying these methods. A key application is the use of centrality measures, which is explored by \citep{crucitti2006centrality}. They investigate the distributions of different centrality measures—including closeness, betweenness, and straightness—in the graphs of several cities. Their finding that these metrics reveal different, non-trivially correlated aspects of network structure provides a valuable precedent for our study, underscoring the necessity of using a combination of measures to gain a comprehensive understanding of a transportation network.

These analytical methods are often applied to address key urban goals like resilience, equity, and optimization. The relationship between equality of access and systemic resilience is explored by \citep{fan2022equality}, who find that a more equitable distribution of facilities enhances a network's tolerance to disruption. To practically quantify robustness, \citep{cordero2022merger} propose a methodology for creating a composite resilience index by merging multiple network graph indicators into a single score. This work is directly relevant as it provides a precedent for using a multi-indicator approach to assess the resilience of San Diego's transportation network. Beyond resilience, \citep{pana2024graph} provide a comprehensive overview of how classical graph theory algorithms can be applied to solve practical optimization problems, justifying our use of specific algorithms to analyze and identify potential improvements within the network.

Finally, our research is situated within a movement towards more advanced and holistic modeling frameworks. Recognizing that urban infrastructures are interdependent, \citep{schoonenberg2019hetero} introduce a Hetero-functional Graph Theory capable of modeling complex interactions between systems like transportation and power grids. To capture temporal changes, research has moved towards dynamic models. \citep{ding2018detecting} propose a multi-layer network methodology, where each layer represents the network's state at a specific time. Extending this, \citep{phua2024fostering} utilize a Dynamic Knowledge Graph (DKG) to encode rich, semantic information that evolves over time, allowing for more nuanced simulations. This trajectory toward data-rich, dynamic modeling culminates in visionary frameworks like MetaCity, introduced by \citep{zhang2025metacity}, which aims to create a "digital twin" of a city for holistic simulation and planning. This work positions our analysis of San Diego's transportation network as a vital component within this larger, cutting-edge pursuit of urban science.
\section{Dataset Description}
This study utilizes several open-source datasets to construct a multi-layered representation of San Diego's transportation network. The primary data sources and processing tools are outlined below.

\subsection{OpenStreetMap (OSM) Data}
The foundational geospatial data for this research is sourced from OpenStreetMaps (OSM) \cite{OpenStreetMap}, a collaborative, open-source project that provides global mapping and infrastructure data. OSM offers a rich, graph-oriented data structure that includes road networks, pedestrian pathways, and Points of Interest (POIs). The analysis is facilitated by the OSMnx Python library (v. 2.0.5), which is built upon NetworkX and specifically designed to download, model, and analyze street networks from the OSM database. For this study, we defined a bounding box encompassing the San Diego metropolitan area and extracted two distinct network layers:
\begin{itemize}
    \item Driving Network: Includes all roads accessible to automobiles.
    \item Walking Network: Includes pedestrian pathways, sidewalks, and trails.
\end{itemize}

Additionally, POI data was extracted from OSM to identify key urban amenities and destinations across various categories (e.g., commercial, recreational, civic).

\subsection{San Diego Public Transit Data}
While OSM provides the locations of public transportation stops, it lacks comprehensive route information. To overcome this limitation, we incorporated public transit data from the San Diego Open Dataset portal. Specifically, we utilized two shapefiles:
\begin{itemize}
    \item The Transit Route Dataset: Contains line geometries representing all public transit routes within the city.
    \item The Stop Dataset: Contains point geometries marking the precise locations of all transit stops.
\end{itemize}

The dataset’s spatial extent, or bounding box, is derived from the union of the minimum and maximum coordinates of these shapefiles. The data was transformed from its projected Coordinate Reference System (CRS) to the WGS84 (EPSG:4326) geographic coordinate system to ensure compatibility with the OSM data. This combined dataset provides a detailed representation of public transit infrastructure, enabling a more nuanced analysis of accessibility and multi-modal connectivity.

\section{Observations and Analysis}
This section details our multi-faceted analysis of San Diego's transportation network. We begin by outlining key methodological considerations for spatial network analysis, then dissect the network's structure through Points of Interest and centrality measures, and conclude by identifying functional mobility patterns using community detection. Together, these analyses reveal a system defined by a stark core-periphery divide, modal conflicts, and critical infrastructure vulnerabilities.

\subsection{Methodological Considerations and Preliminary Observations}
Initial analysis of the full San Diego metropolitan area revealed that the high density of nodes and edges rendered fine-grained observation of local network interactions difficult. Furthermore, centrality calculations on the full graph proved to be computationally intensive. To address these challenges while enabling detailed local analysis, the analytical scope was narrowed to a 5,000-meter radius centered on the University of California, San Diego (UCSD). This context window provides a more manageable yet meaningful subgraph that encompasses the university campus, the surrounding La Jolla community, and major transportation corridors, allowing for a more effective visualization and analysis of node-edge relationships. This focused approach enabled the generation of distinct walking and driving network layers for direct comparison.

\begin{figure}[ht]
    \centering
    \includegraphics[width=0.2\textwidth]{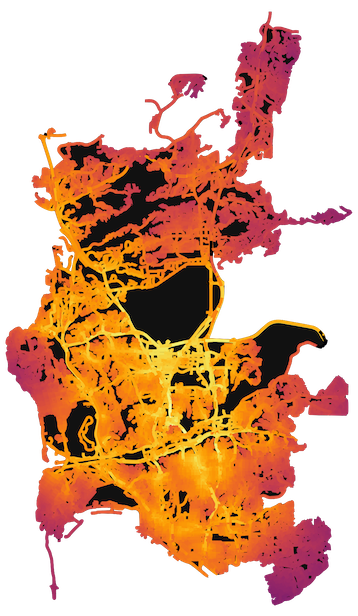}
    \caption{Closeness centrality show on San Diego. Brighter colors show more central edges.}
    \label{fig:CC_heatmap}
\end{figure}

The initial analysis of the focused subgraph revealed several important methodological considerations and preliminary findings. Centrality calculations, visualized as a heatmap (Figure~\ref{fig:CC_heatmap}), immediately highlighted major freeways and arterial roads as structurally important, confirming the viability of the analytical approach. However, this process also surfaced two critical caveats inherent to this type of spatial network analysis.

First, a notable ``edge effect'' was observed as seen in Figure~\ref{fig:CC_heatmap}, where centrality values for nodes near the boundary of the selected context window appear artificially suppressed relative to more central nodes. This phenomenon occurs because centrality is calculated relative to the subgraph, not the global network, limiting the number of paths that can traverse nodes at the periphery. This boundary effect is an intrinsic property of analyzing any bounded subgraph and must be considered when interpreting results, particularly near the edges of the study area.

Second, the analysis identified a large area of anomalous low connectivity corresponding to the Marine Corps Air Station Miramar. The absence of road and pathway data in this zone suggests that information for sensitive locations, such as military installations, has been systematically redacted from the public OSM dataset. This observation underscores the importance of acknowledging that publicly sourced data may not be fully transparent or complete. Such intentional data fragmentation can influence the calculated network properties of adjacent areas and must be accounted for in the analysis.

\begin{figure}[ht]
    \centering
    \begin{subfigure}[c]{0.4\textwidth}
        \includegraphics[width=\textwidth]{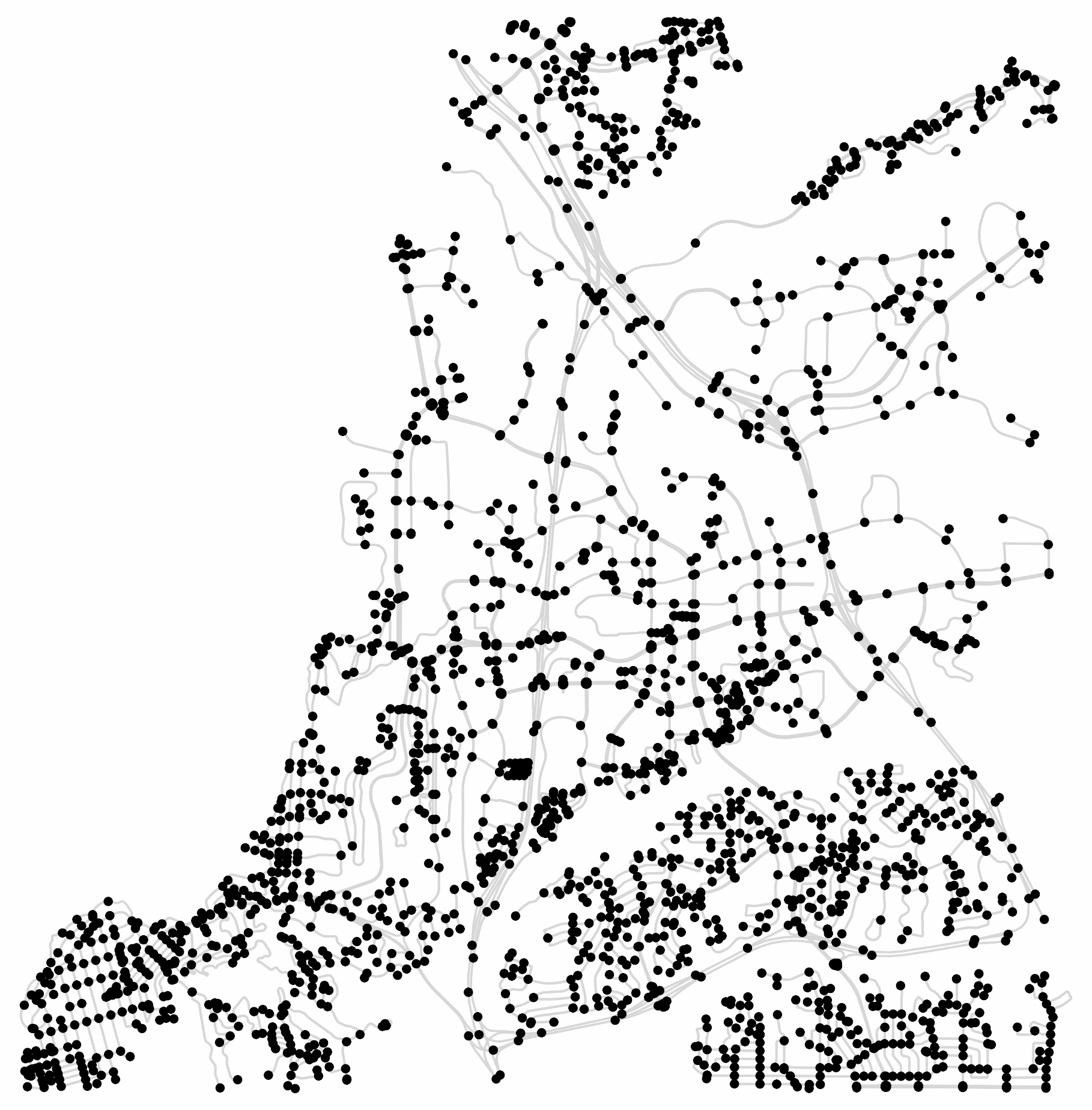}
            \caption{UC San Diego area driving graph}
        \label{fig:G_drive}
    \end{subfigure}
    \hspace{0.2in}
    \begin{subfigure}[c]{0.4\textwidth}
        \includegraphics[width=\textwidth]{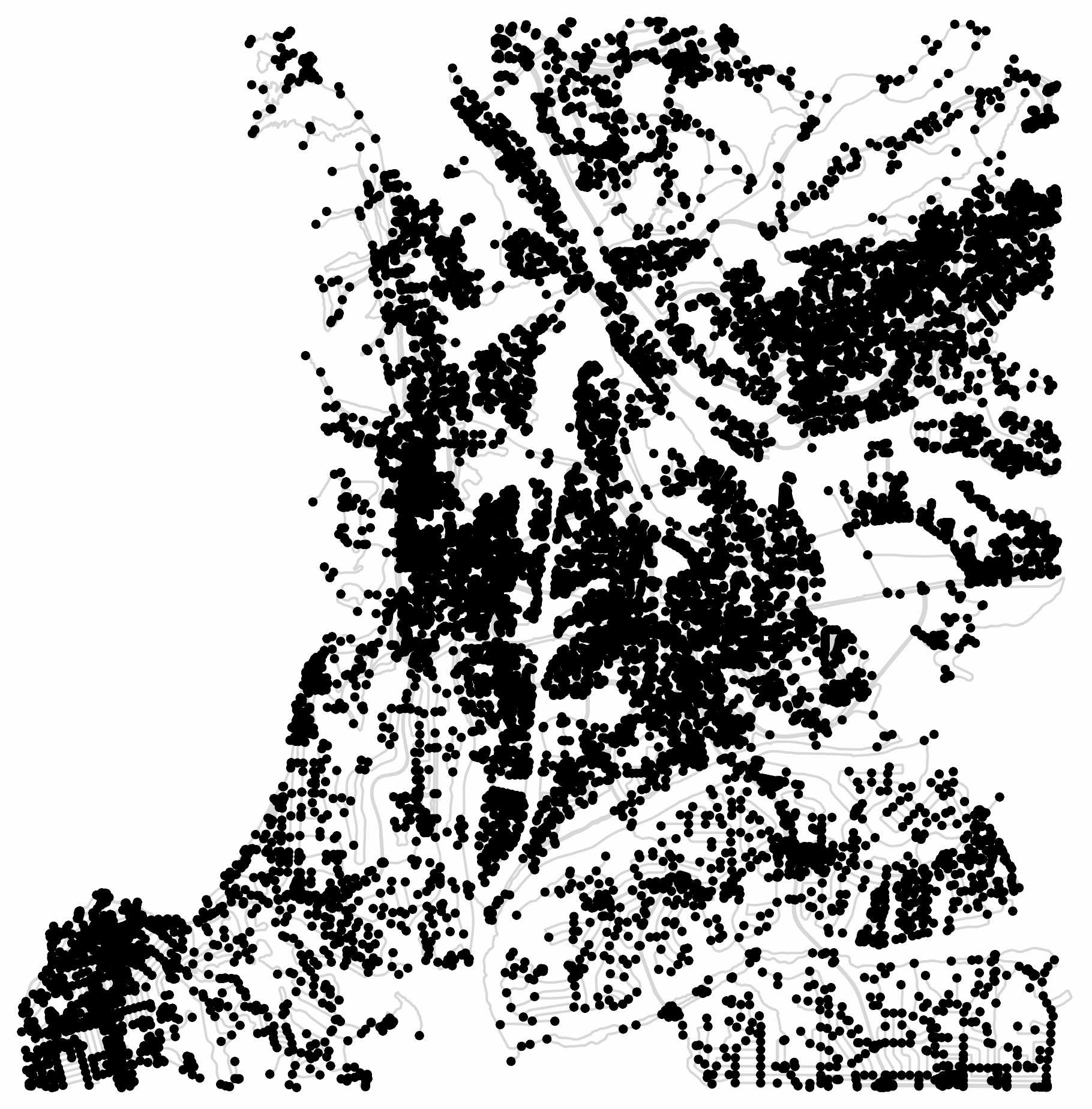}
            \caption{UC San Diego area walking graph}
        \label{fig:G_walk}
    \end{subfigure}
    \caption{Visualizations of different layers (driving on the left and walking on the right) for UC San Diego area}
    \vspace{-0.1in}
    \label{fig:G_walk_drive}
\end{figure}

A direct comparison of the driving (Figure~\ref{fig:G_drive}) and walking (Figure~\ref{fig:G_walk}) networks within the UCSD-centered context revealed significant structural differences. While there is considerable geographic overlap between the two networks, the walking graph is visibly denser, containing more nodes and edges within the same area.

Furthermore, the comparison highlights a distinct inverse relationship between the two modalities. Major freeways (I-5, I-805, SR-52), which emerge as high-centrality corridors in the driving graph, simultaneously function as prominent barriers or boundaries in the walking graph, exhibiting low connectivity for pedestrians. Conversely, some areas that appear as barriers in the driving network, such as canyons and ridges, are found to contain dedicated walking paths, indicating a unique pedestrian infrastructure. These findings confirm that the separate modal layers are independently rich with contextually meaningful data, validating the layered approach and setting the stage for a more detailed multi-modal analysis.

\subsection{Points of Interest and Public Transit Analysis}
Integrating Points of Interest (POIs) with the public transportation network reveals the functional structure of urban accessibility and its underlying inequities. Our analysis focuses on two key themes: the network's core structure and the resulting gaps in access.

\subsubsection{Assessing Network Structure and Transit Accessibility}
The integrated multilayer network consists of 22,156 total nodes, including 15,996 POIs and 6,160 transit stops. A defining characteristic of this network is its edge distribution: of 11,472 total edges, 11,155 are inter-layer edges connecting POIs to transit stops, while only 317 are intra-transit edges connecting stops along routes. This dramatic imbalance leads to a critical insight about the system's design:
\begin{observation}\label{obs:poi-driven}
    The network’s connectivity is primarily driven by POI-to-stop linkages rather than extensive route interconnectivity. 
\end{observation}

\begin{figure}[ht]
    \centering
    \begin{subfigure}[c]{0.45\textwidth}
        \includegraphics[width=\textwidth]{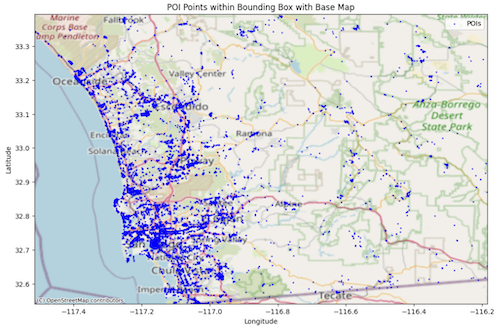}
            \caption{Points of interest visualization}
        \label{fig:poi}
    \end{subfigure}
    \begin{subfigure}[c]{0.45\textwidth}
        \includegraphics[width=\textwidth]{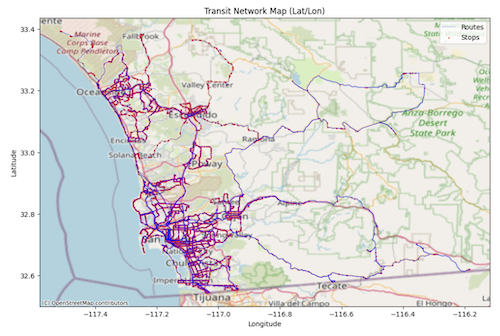}
            \caption{San Diego transit routes}
        \label{fig:transit}
    \end{subfigure}
    \caption{Visualizations of (a) points of interest and (b) transit routes of San Diego.}
    \vspace{-0.1in}
    \label{fig:poi_trnasit}
\end{figure}

\begin{figure}[ht]
    \centering
    \begin{subfigure}[c]{0.45\textwidth}
        \includegraphics[width=\textwidth]{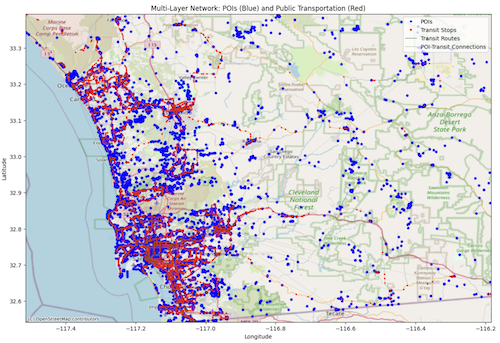}
            \caption{Two dimensional visualization of network layers}
        \label{fig:2d}
    \end{subfigure}
    \begin{subfigure}[c]{0.45\textwidth}
        \includegraphics[width=\textwidth]{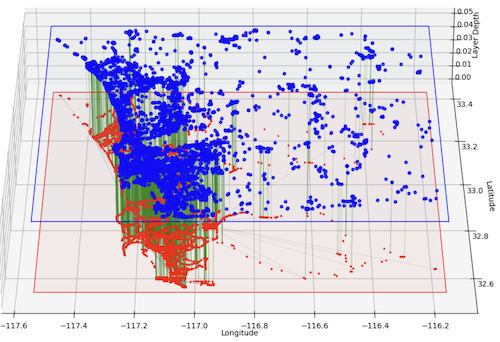}
            \caption{Three dimensional visualization of network layers}
        \label{fig:3d}
    \end{subfigure}
    \caption{Visualizations of multilayer network in (a) 2D and (b) 3D}
    \vspace{-0.1in}
    \label{fig:2d_3d}
\end{figure}

This structure suggests a hub-and-spoke model where accessibility is defined by proximity to individual stops, rather than by a highly interconnected web of transit routes. While this design connects 69.7\% of POIs to transit within a 500-meter radius , the network's overall integration is limited, as reflected by a very low average degree centrality of 0.0006. The visualizations in Figure~\ref{fig:poi_trnasit} and Figure~\ref{fig:2d_3d} illustrate this structure, showing dense clusters of connectivity in urban cores but sparse linkages elsewhere.

The following figures visually substantiate this network structure and its spatial distribution across San Diego County. Figure~\ref{fig:poi_trnasit} illustrates the core-periphery divide: (a) the dense clustering of Points of Interest (POIs) in urban centers (b) mirrors the concentration of the transit network's routes and stops in the same areas, while both become sparse in rural regions. Furthermore, Figure~\ref{fig:2d_3d} provides a 2D and 3D rendering of the integrated multilayer network, where the inter-layer connections (green lines) make the accessibility gap explicit, showing a dense web of connections in the downtown hub and visually isolated POIs in outlying areas.

The multilayer representations in Figure~\ref{fig:2d_3d} make this accessibility gap explicit by visualizing the inter-layer connections between POIs and the transit system. In these figures, POIs (blue dots) are linked to nearby transit stops (red dots) by green lines if they are within a 500-meter radius. The resulting image confirms the insights from the quantitative data: a high density of these connections forms a robust network hub in downtown San Diego, where many POIs are accessible. However, in outlying coastal and eastern areas, these connections become sparse or non-existent, visually affirming that a significant portion of POIs are isolated from the transit system. This aligns with the data in Table~\ref{table:summary}, which quantifies the extent of this divide. 

\begin{table}[ht]
    \caption{Summary of Multilayer Network Analysis Results for San Diego County}\label{table:summary}
    \begin{center}
        \begin{tabular}{ll}
         \textbf{Category} &\textbf{Value} \\
        \hline \\[-4.8pt]
        \textbf{Network Statistics} & \\
        Total Nodes & 22,156 (POIs: 15,996) \\
        Total Edges & 11,472 (Inter-Layer: 11,155) \\
        Average Degree Centrality  & 0.0006 \\
        Maximum Degree Centrality  & 0.0072 \\
        POIs Connected & 11,155 \\
        POIs Isolated (> 500m from stops) & 4,841\\
        \hline \\[-4.8pt]
        \textbf{Spatial Distribution Insights} & \\
        High Centrality Nodes (> 1.5x average) & 4,432\\
        High Centrality Nodes in Downtown & 362\\
        \hline \\[-4.8pt]
        \textbf{Connectivity and Accessibility} & \\
        Percentage of POIs Connected to Transit  & 69.7\% \\
        Average Distance of Inter-Layer Connections (meters) & 89.9
        \end{tabular}
    \end{center}
\end{table}

\subsubsection{Identifying Equity Gaps in the Core-Periphery Divide}
While the network provides substantial coverage in urban centers, our analysis quantifies a significant core-periphery divide. The most telling statistic is that 4,841 POIs (30.3\%) are isolated, located more than 500 meters from the nearest transit stop.
\begin{observation}
    The analysis highlights a well-integrated urban core with significant accessibility, but the 30.3\% isolated POIs suggest potential equity issues in rural access.
\end{observation}

The spatial analysis of degree centrality provides quantitative evidence for this core-periphery structure. Of the 4,432 nodes identified as having high connectivity (degree centrality > 1.5x the average), a disproportionate 362 nodes—or approximately 8.2\%—are concentrated within Downtown San Diego. This intense concentration reflects a highly integrated urban core where, as seen in Figure~\ref{fig:2d}, numerous POIs and transit routes are densely interwoven. While the remaining high-centrality nodes indicate additional smaller hubs along coastal and eastern corridors like Oceanside and El Cajon (visible in Figure~\ref{fig:transit}), the dominance of the downtown core is a key structural feature of the network. This finding is further supported by the dense central layer in the 3D visualization (Figure~\ref{fig:3d}), which visually confirms the network's urban-centric design.

\subsection{Network Attributes Analysis}
Beyond overall structure, we analyzed specific network attributes to understand route efficiency, network resilience, and walkability.

\subsubsection{Shortest Path}
\begin{figure}[ht]
    \centering
    \begin{subfigure}[c]{0.45\textwidth}
        \includegraphics[width=\textwidth]{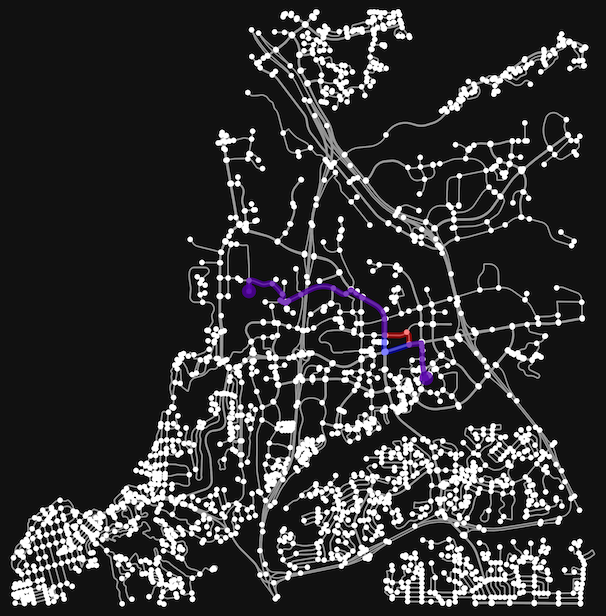}
            \caption{Driving route: Shortest distance (red), shortest duration (blue)}
        \label{fig:d_shortest}
    \end{subfigure}
    \begin{subfigure}[c]{0.45\textwidth}
        \includegraphics[width=\textwidth]{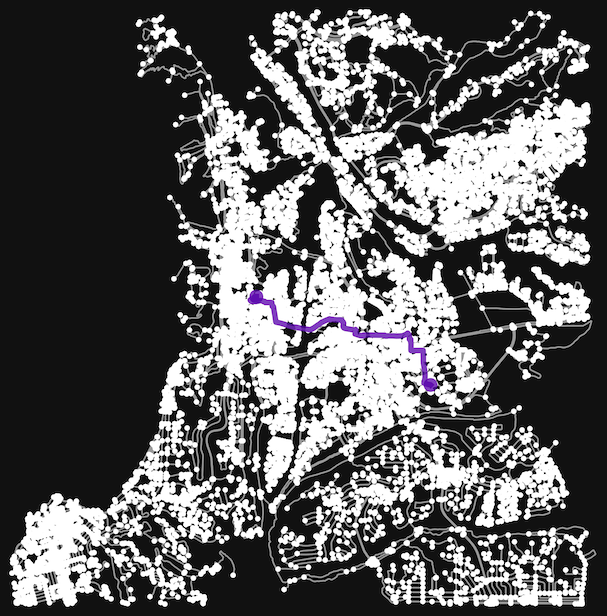}
            \caption{Walking route visualization: Shortest distance (red), shortest duration (blue)}
        \label{fig:w_shortest}
    \end{subfigure}
    \caption{Visualizations of different optimization in (a) driving and (b) walking layers}
    \vspace{-0.1in}
    \label{fig:drive_walk_min}
\end{figure}

We implemented Dijkstra’s algorithm \cite{dijkstra2022note} to calculate shortest paths based on both distance and travel time for driving and walking layers (Figure~\ref{fig:drive_walk_min}). The primary purpose of this analysis was to demonstrate the flexibility of our modeling framework:
\begin{observation}
    This framework enables us to optimize any objective (e.g. least number of turns, distance, time) algorithm on any (multiplex) network. 
\end{observation}
The model can be augmented to include variables like intersection delays or altitude changes, making it a robust tool for dynamic route optimization. We were able to adjust attributes per our edges to associate to speeds for a shortest travel-time calculation as seen in Figure~\ref{fig:drive_walk_min}. 

We used the OSMnx library2, which wraps NetworkX3 shortest path, which itself is an implementation of Djikstra’s Algorithm. With this framework, we determined that we could also use the same procedure for other forms of optimizations in route-selection, and craft a dynamic function call to recursively map value-attributes per edge and utilize the same shortest path function. We have mapped two points on our local graph between Geisel Library to Chicago Fire Grill (burgers) (Figure~\ref{fig:G_walk}). The OSMnx library uses API calls to Geocoding service Nominatim 5 to resolve place names to geographical coordinates in Latitude/Longitude. Our framework not only enables us to include augmentations but also we can use multiple transportation modes to optimize the shortest distance (e.g. time or distance) function on a multiplex transportation network. 

\subsubsection{Centrality Analysis: Uncovering Walkability and Resilience}
In this section, using two centrality measures, we define \emph{walkability} of a certain sub-network, or the whole network, in hand. 

We successfully mapped graph centrality calculations to graph-maps by injecting a color schema as a heatmap per value we calculate per edge, and by projecting that heatmap back into our graph, we can intuitively read our graph and reference the effects. We used a graph-inversion technique to project centrality calculations onto the graph-maps. This technique sets $edge\to node$ and $node\to edge$ and, after calculating, we assign the values to the graph’s edge attributes, and map them to each edge as a heatmap color scheme.

\begin{figure}[ht]
    \centering
    \begin{subfigure}[c]{0.45\textwidth}
        \includegraphics[width=\textwidth]{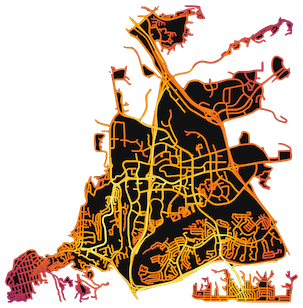}
            \caption{Closeness centrality for driving}
        \label{fig:cc_d}
    \end{subfigure}
    \begin{subfigure}[c]{0.45\textwidth}
        \includegraphics[width=\textwidth]{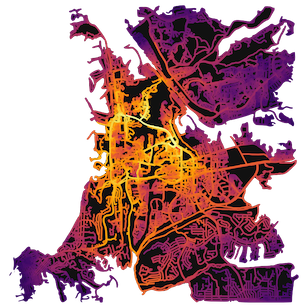}
            \caption{Closeness centrality for walking}
        \label{fig:cc_w}
    \end{subfigure}
    \caption{Visualizations of closeness centrality for two layers}
    \vspace{-0.1in}
    \label{fig:cc_drive_walk}
\end{figure}

\paragraph{\textbf{\textit{Closeness}}}
We calculated our Closeness centrality \footnote{$C(x) = \frac{N-1}{\sum_y d(x, y)}$ where $d(y,x)$ is the distance (length of the shortest path) between vertices $x$ and $y$ and $N$ is the number of nodes in the graph.} for our two graphs and the differences are very apparent. Since the entire intent and purpose of roadways is the same as the metric of closeness centrality, in Figure~\ref{fig:cc_d} we can see here that the entire mapping holds high values for closeness. This tells us that the infrastructure of our roadways are designed well with this in mind. The walking graph in Figure~\ref{fig:cc_w} tells a different story, where the highlighted areas are somewhat isolated to the grouping within UCSD, UCSD’s housing area, with the inclusion of UTC Mall and the strip malls around Nobel Drive all holding a (relatively) walkable area. Outside of these areas, walkability is extremely limited or isolated. 
\begin{observation}
    Closeness centrality analysis gives us multiple insights:
    \begin{itemize}
        \item Closeness centrality gives us a good understanding of the efficiency of the design;
        \item San Diego, generally speaking, is not a walkable city. 
    \end{itemize}
\end{observation}

This is a bitter observation. As we have heard before that our cities are designed with car in mind and are not human-centered, the network science metrics agree with that. 

\begin{figure}[ht]
    \centering
    \begin{subfigure}[c]{0.45\textwidth}
        \includegraphics[width=\textwidth]{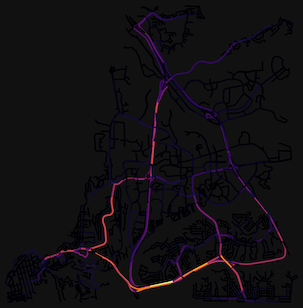}
            \caption{Betweenness centrality for driving}
        \label{fig:bc_d}
    \end{subfigure}
    \begin{subfigure}[c]{0.45\textwidth}
        \includegraphics[width=\textwidth]{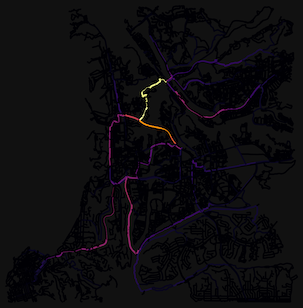}
            \caption{Betweenness centrality for walking}
        \label{fig:bc_w}
    \end{subfigure}
    \caption{Visualizations of betweenness centrality for two layers}
    \vspace{-0.1in}
    \label{fig:bc_drive_walk}
\end{figure}

\paragraph{\textbf{\textit{Betweenness}}}
For our betweenness centrality\footnote{$g(v) = \sum_{s\neq v\neq t}\frac{\sigma_{st}(v)}{\sigma_{st}}$ where $\sigma _{st}$ is the total number of shortest paths from node $s$ to node $t$ and $\sigma _{st}(v)$ is the number of those paths that pass through $v$ (not where $v$ is an end point).}, in Figure~\ref{fig:bc_drive_walk} we can see highlighted are effectively transit bottlenecks. For our Driving graph shown in Figure~\ref{fig:bc_d}, this demonstrates that the freeways have become essential for any flow of traffic in or around the city, with almost no values for alternative pathways assigned outside of freeways. This extreme reliance on a few corridors indicates low network redundancy and a significant resilience problem.
\begin{observation}
    The freeways are almost too critical with little to none escape routes. 
\end{observation}

For the Walking graph in Figure~\ref{fig:bc_w}, we can see a little more localized effects demonstrated. Contextually, we can see the highlighted areas are walking pathways leading out of UCSD to the northeast and southeast. Since UCSD has good walkability internally, the bottlenecks are aligned with crossing major road-throughways: namely crossing over the I-5 freeway and out of the boundaries of UCSD. The second most highlighted areas in the walkability bottlenecks graph are major roadways, as they have sidewalks that span the distances between major areas. This is pretty straightforward, since almost no consideration for walking infrastructure is generally considered beyond just having main street sidewalks, unless a zone depends on foot traffic. 

\subsubsection{A Proposed Metric: The Network-based Walkability Score}
To quantify the insights from our centrality analysis, we propose a novel metric to score the walkability of a given area. This score synthesizes the concepts of efficiency (closeness) and choke points (betweenness).

\begin{definition}
    The \emph{walkability score} of an area is as follows:
    \begin{equation}
        W = \frac{\sum_{i\in A} C_i}{\sum_{i\in A}B_i}
    \end{equation}
    where $A$ is the given area, $C_i$ is the closeness centrality of a path, and $B_i$ is the betweenness centrality of a path. 
\end{definition}

\subsection{Communities Analysis}
Transportation networks consist of multiple modes of travel, including walking, driving, biking, bus, and train. By improving our understanding of structural organization of these networks, we can attempt to optimize urban mobility, which would prove to be useful in city planning situations. Community detection techniques, which help identify connected groups of nodes, may help reveal functional local transport community structures of highly accessible regions. In this study, we aim to analyze transportation networks within San Diego using OSM data. We model these networks as multilayer graphs, where each layer corresponds to a different transportation mode. By applying community detection, we can investigate how different modalities bring rise to different localized transport clusters. Multilayer community detection has been previously applied to transportation systems to  capture deep intricacies in transportation networks that are unable to be captured by simple single layer methods. \citep{huang2021survey} surveys multilayer community detection methods across multiple papers, demonstrating how communities in walking, bus, and driving layers exhibit distinct topological properties. Both multiplex and multilayer analyses are discussed in this paper; multiplex networks being graphs where nodes are shared across layers but with different attributes, and where multilayer networks permit differing nodes but must have  interlayer edges. Though multilayer community detection methods \cite{alessandretti2023multimodal} are less studied, they better represent real world transportation systems (but due to limitations in OSM may not be fully applied in our study).

\citep{alessandretti2023multimodal} describes a multilayer analysis to identify community structures in the Brisbane Metropolitan Area using \citep{blondel2008fast}’s algorithm, more commonly known as the Louvain community detection method. Given data sources of bus GPS trajectories, passenger smart card transactions, and roadside Bluetooth vehicle detections for car trajectories, each layer is jointly combined and data points are grouped into Voronoi cells with a 1 KM radius. Trajectories of flows between cell centers (seeds) are mapped as edges in the dual (unsure if this is the right term) to form a new graph. Each layer is then split up and individually have Louvain community detection
applied. Findings from this study show differing structure of communities formed between the layers, with a novel comparison technique being used to describe similarity of trajectories of each layer. The car layer has the greatest dissimilarity to the other layers, to which point the authors hint that car users are less bound by the overall mono-centric structure of the city. A primary difference from this study would be that it investigates flow, while we only investigate travel time.

We begin by constructing a set of graphs fetched using OSMNX, a Python package which retrieves data from OpenStreetMap and generates spatial graphs. Our model treats these basic transportation networks as weighted graphs by using OSMNX functions to calculate travel times for each edge, which represents a road or transportation method’s segment (driving road, biking path, walking path). In order to complete calculations within a reasonable time constraint, we define our region of study as a 5-kilometer radius centered around UCSD. Using the OSMNX library, we extract three separate pre-defined layers, each corresponding to a distinct mode of transportation: ``drive'', ``walk'', ``bike''. Ideally, we would be able to retrieve bus routes as well but due to limitations of the OSMNX library and OSM itself (which defines bus routes as polygonal relations and not as tags on road segments themselves) we are not able to easily apply this analysis. Each of these networks are constructed as a MultiDirectionalGraph – these are directed graphs which allow for self-loops, according to the NetworkX library. Each vertex represents OSM nodes (intersections and endpoints), with each edge representing path segments: we augment these path segments by using OSMNX functions to provide travel speed and travel time for each edge. To model community structures for each network, we used the Louvain method for community detection, already implemented within NetworkX. The Louvain method for community detection is a greedy optimization method optimizing for modularity as described in \citep{blondel2008fast}. Modularity is a measure of relative density of edges inside versus outside of a community – it can be seen as a measure of the quality of a community. We are able to provide a weight parameter and a resolution parameter – we use travel time as our weight and intend to investigate how community formation differs as we vary the resolution parameter. It is known that \emph{smaller values of resolution favor constructions of larger communities, while larger values of resolution should create an increased amount of smaller communities.} 

We retrieved bus information from MTS. Data retrieved is stored in a JSON – it provides stop latitude, longitude, and bus arrival time to the nearest minute. By retrieving each bus route individually for each route, we can create a network of bus stops with travel times determined by time deltas between each bus stop arrival time. To combat the issue of time deltas being zero for nearby stops, we will assume that the minimum travel time for the bus is 30 seconds. 

\begin{figure}[ht]
    \centering
    \begin{subfigure}[c]{0.3\textwidth}
        \includegraphics[width=\textwidth]{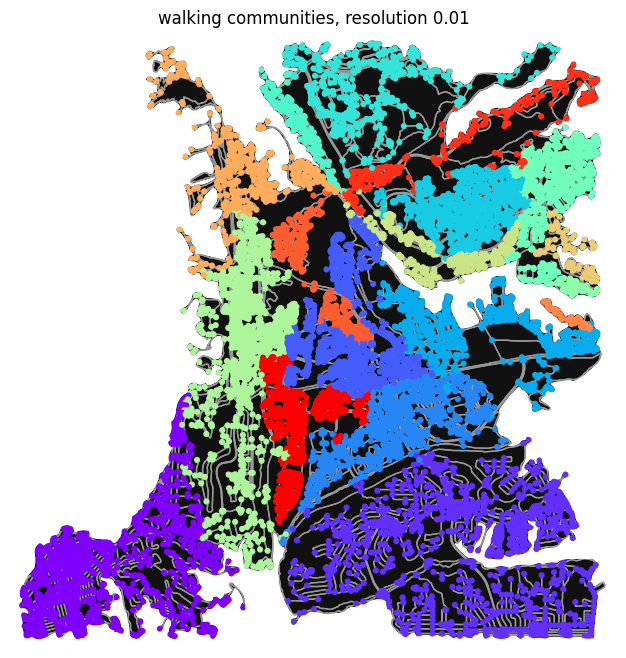}
            \caption{Resolution = 0.01}
        \label{fig:res_01}
    \end{subfigure}
    \begin{subfigure}[c]{0.3\textwidth}
        \includegraphics[width=\textwidth]{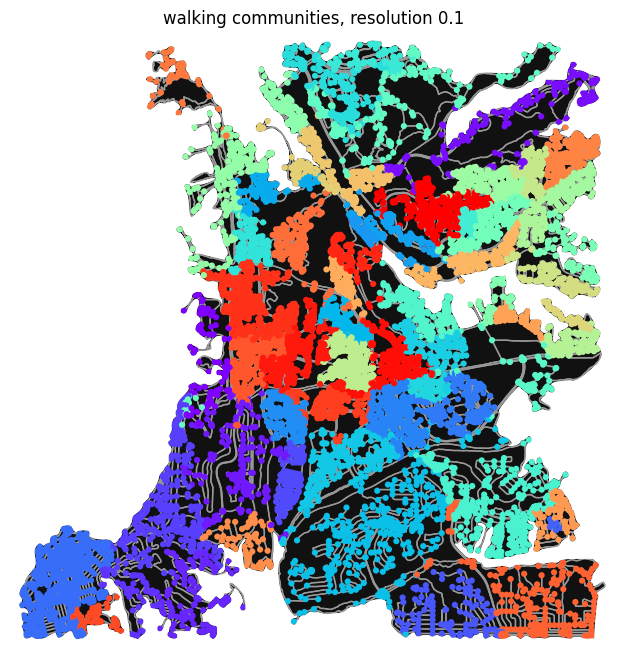}
            \caption{Resolution = 0.1}
        \label{fig:res_1}
    \end{subfigure}
    \begin{subfigure}[c]{0.3\textwidth}
        \includegraphics[width=\textwidth]{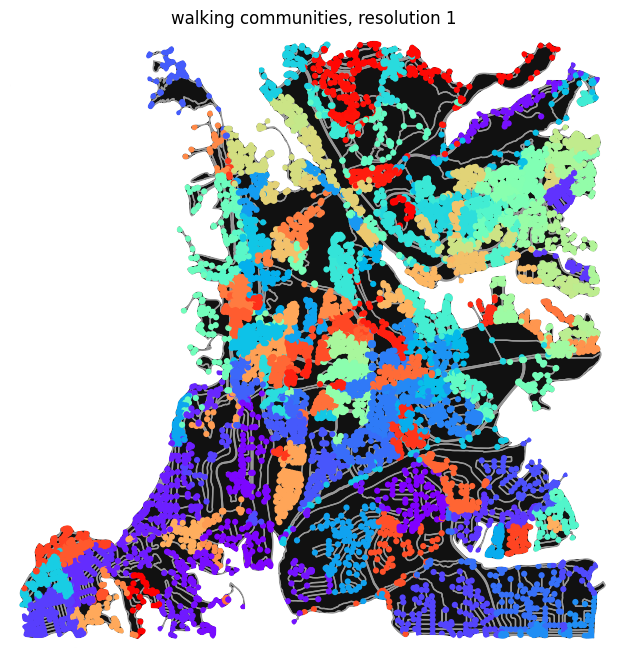}
            \caption{Resolution = 1}
        \label{fig:res_10}
    \end{subfigure}
    \caption{Communities with resolution (a) 0.01, (b) 0.1, and (c) 1}
    \vspace{-0.1in}
    \label{fig:communities}
\end{figure}

\begin{figure}[ht]
    \centering
    \begin{subfigure}[c]{0.24\textwidth}
        \includegraphics[width=\textwidth]{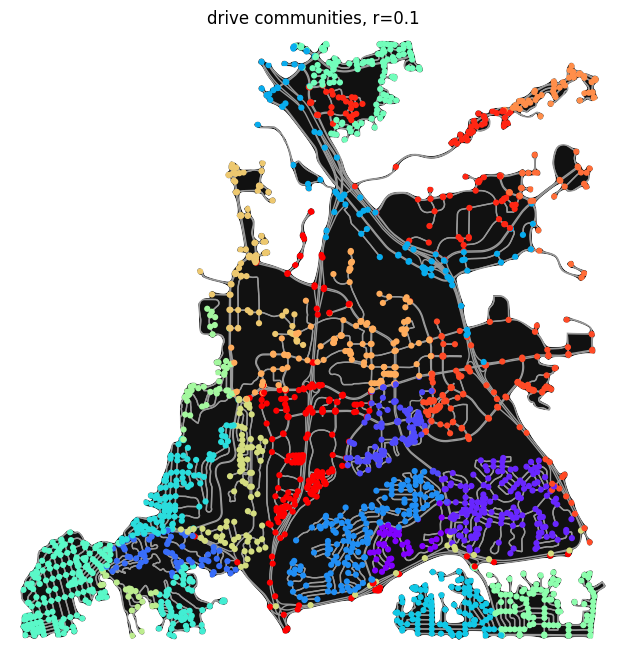}
            \caption{Driving communities}
        \label{fig:com_drive}
    \end{subfigure}
    \begin{subfigure}[c]{0.24\textwidth}
        \includegraphics[width=\textwidth]{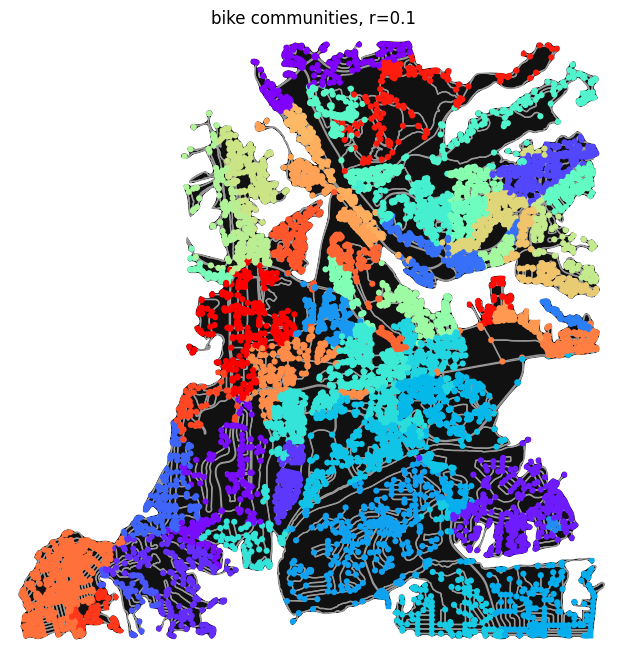}
            \caption{Biking communities}
        \label{fig:com_bike}
    \end{subfigure}
    \begin{subfigure}[c]{0.24\textwidth}
        \includegraphics[width=\textwidth]{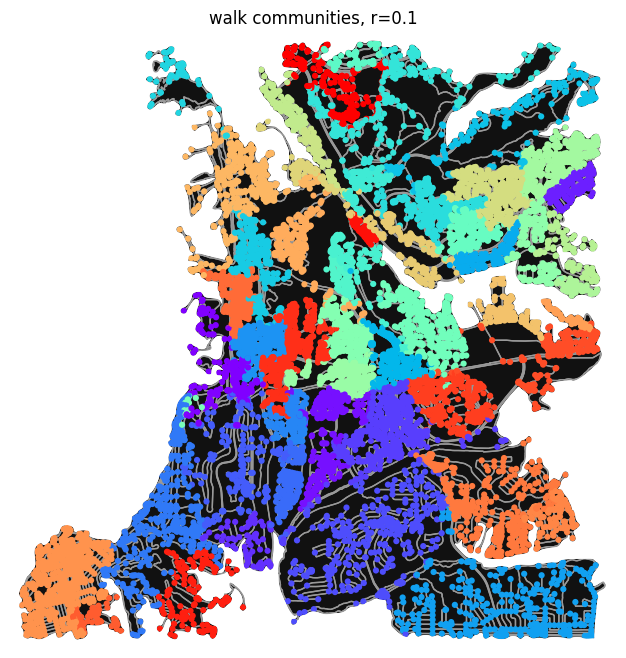}
            \caption{Walking communities}
        \label{fig:com_walk}
    \end{subfigure}
    \begin{subfigure}[c]{0.24\textwidth}
        \includegraphics[width=\textwidth]{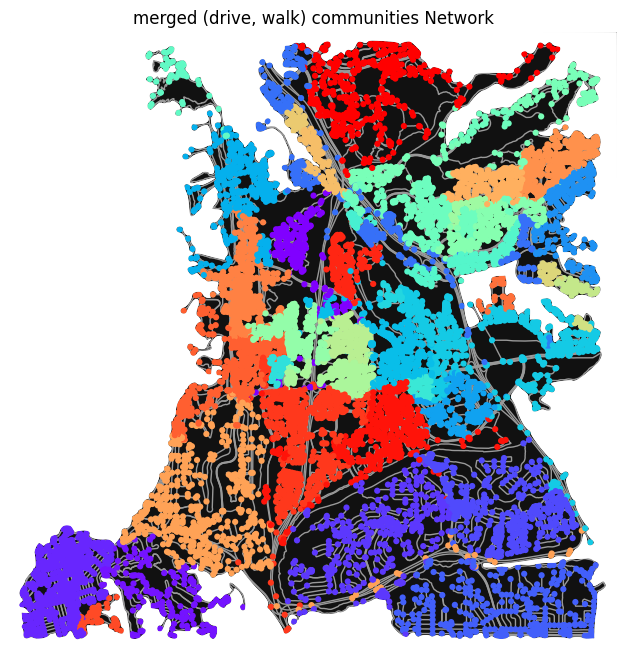}
            \caption{Merged driving and walking communities}
        \label{fig:com_merge}
    \end{subfigure}
    \caption{Communities for different modes of transportation all with resolution 0.1}
    \vspace{-0.1in}
    \label{fig:com}
\end{figure}

A resolution value of 0.1 as seen in Figure~\ref{fig:res_1} gives us the best visibility for our results and reasonable traversal times for communities. Using this resolution parameter, we will generate communities for driving (Figure~\ref{fig:com_drive}) and cycling (Figure~\ref{fig:com_bike}) modalities. We see in our analysis that the faster the modality the larger the community spans in terms of physical area. Note that this may also be a result of greater density of nodes and edges in walking network compared to the driving network. To explore inter-modal transport between nodes, we constructed a combined network from driving and walking, with priority given to walking – essentially, all the nodes and edges of the graphs are combined, with travel time attribute being set to that of the walking graph if there is a conflict. This means that if the user is able to walk, they will walk the path instead of driving, favoring pedestrian accessibility. We apply Louvain community detection with resolution = 0.1 to this composite network, showing us how multi-modal communities may be formed in Figure~\ref{fig:com_merge}: Observe that community distances are larger than in the walking-only network, which suggests that combining transportation modalities gives broader connectivity. The application of Louvain community detection on our mutlilayer graph construction reveals that transport network structure is highly sensitive to resolution parameters and the type of transport. Walking networks will produce more spatially compact, smaller, and more communities than that of driving networks, which should cover broader areas. The combination of modalities significantly changes the community landscape. We also look to test community detection on our bus routes. To merge the multiple layers of bus routes together, we set a threshold of physical distance. It is assumed that passengers will be willing to walk from one stop to another under a certain radius threshold – we set this to 200 meters in our analysis, with the expected pedestrian walking speed to be 1.4 m/s. Bus stop nodes will join together with an edge if their distance is under 200 meters, with the travel time weight set to $distance/1.4$.

\section{Conclusion and Future Works}
This paper demonstrates the utility of network science for modeling and analyzing a real-world, multi-modal transportation system in San Diego. By applying centrality analysis, community detection, and multi-layer network visualization, our study has uncovered key structural characteristics, functional dependencies, and systemic vulnerabilities that offer valuable insights for urban planning and transportation equity.

Our analysis reveals a stark dichotomy between the city's driving and walking networks. The driving layer is characterized by high connectivity across a broad area, but this efficiency is precarious; we observe that the network is overly reliant on critical freeways with few alternative routes, indicating a lack of resilience. Conversely, the walking network is highly fragmented, with high closeness centrality confined to dense, isolated pockets, confirming that San Diego is not a broadly walkable city. Furthermore, our multi-layer analysis of the public transit system shows that while the urban core is well-integrated, connectivity is primarily driven by direct Point of Interest (POI) to transit-stop linkages rather than by a deeply interconnected web of routes. This structure contributes to significant accessibility gaps, with our findings highlighting that 30.3\% of POIs remain isolated, pointing to potential equity issues for residents in peripheral and rural areas.

Ultimately, this work provides a flexible and powerful framework capable of optimizing for various objectives—such as time, distance, or complexity—on multimodal transportation networks. It effectively transforms raw geospatial data into an actionable model that identifies systemic strengths, critical vulnerabilities, and clear disparities in urban mobility.

The primary limitations of this study are rooted in the nature of real-world, open-source data. The OpenStreetMap (OSM) and transit data, while extensive, suffer from inconsistent labeling standards and missing information. During our analysis, we encountered challenges that necessitated data manipulation—for instance, manually assigning realistic speeds to walking paths where default values were inappropriate. While such interventions are common in practical data science, they represent a potential source of imprecision. Acknowledging these data imperfections is crucial, as they inherently limit the granularity and absolute accuracy of the model's current predictions.

The findings and limitations of this study open several promising avenues for future research aimed at refining the model and translating its insights into practical applications.

First, the model's fidelity can be enhanced by incorporating more sophisticated, weighted network measures. For example, using a weighted centrality metric like PageRank that accounts for travel time, road capacity, or public transit frequency could provide a more dynamic and realistic assessment of network flow and congestion, particularly regarding the critical freeway corridors we identified. Similarly, integrating real-time transit feeds (e.g., GTFS-Realtime) would allow for a dynamic analysis that captures service disruptions and fluctuating travel times.

Second, future work should focus on direct policy and planning applications. We propose leveraging this framework to collaborate with agencies like the San Diego Metropolitan Transit System (MTS). The model can be used to: (1) Optimize stop placement to strategically reduce the percentage of isolated POIs and improve service to underserved communities. (2) Simulate the impact of new bus routes or infrastructure projects, allowing planners to test interventions and prioritize those that most effectively enhance both equity and network resilience. (3) Inform equitable infrastructure expansion by using centrality and accessibility heatmaps to pinpoint and target connectivity gaps in peripheral and rural areas.

By pursuing these directions, this research can evolve from a powerful diagnostic tool into a proactive planning instrument that helps build a more resilient, efficient, and equitable transportation system for San Diego.

\section*{Acknowledgements}
We would like to thank the instructor of ECE 227 course Massimo Franceschetti for his guidance through his course while this paper was a project in the course. 

\bibliography{trb_template}


\end{document}